\documentstyle[emulateapj]{article}

\def\lsim{\mathrel{\rlap{\lower 4pt \hbox{\hskip 1pt $\sim$}}\raise 1pt
\hbox
        {$<$}}}
\def\gsim{\mathrel{\rlap{\lower 4pt \hbox{\hskip 1pt $\sim$}}\raise 1pt
\hbox
        {$>$}}}

\newcommand{\kms}{km~s$^{-1}$}

\lefthead{Nakamura et al.}
\righthead{Light Curve and Spectral Models for the Hypernova SN 1998bw
associated with GRB980425}

\begin{document}

\submitted{Accepted for publication in ApJ (12 November 2000)}

\title{Light Curve and Spectral Models for the Hypernova SN 1998bw
associated with GRB980425}

\author{Takayoshi Nakamura$^{1}$, Paolo A. Mazzali$^{2,5}$,
Ken'ichi Nomoto$^{3,5}$, Koichi Iwamoto$^4$}

\altaffiltext{1}{Department of Astronomy,
School of Science, University of Tokyo, Tokyo, Japan;
nakamura@astron.s.u-tokyo.ac.jp}

\altaffiltext{2}{Osservatorio Astronomico di Trieste, Trieste, Italy;
mazzali@ts.astro.it}

\altaffiltext{3}{Department of Astronomy,
School of Science, University of Tokyo, Tokyo, Japan;
nomoto@astron.s.u-tokyo.ac.jp}

\altaffiltext{4}{Department of Physics, College of Science and Technology,
Nihon University, Tokyo, Japan;
iwamoto@etoile.phys.cst.nihon-u.ac.jp}

\altaffiltext{5}{Research Center for the Early Universe,
School of Science, University of Tokyo, Tokyo, Japan}

\begin{abstract}

A refined model for the unusual Type Ic supernova 1998bw, discovered
as the optical counterpart of GRB980425, is presented, and synthetic
light curves and spectra are compared with the observations. The first
30 days of the light curve and the broad line features of the spectra
can be reproduced with the hydrodynamical model of the explosion of a
14$M_\odot$ C+O star, the core of a star with initial mass
40$M_\odot$, assuming that the explosion was very energetic (kinetic
energy $E_{\rm K} = 5 \times 10^{52}$ erg) and that 0.4$M_\odot$ of
$^{56}$Ni were synthesized.  At late times, however, the observed
light curve tail declines more slowly than this energetic model, and
is in better agreement with a less energetic ($E_{\rm K} = 7 \times
10^{51}$ erg) one. This shift to a less energetic model may imply that
the inner part of the ejecta has higher density and lower velocities
than the model with $E_{\rm K} = 5 \times 10^{52}$ erg, so that
$\gamma$-rays deposit more efficiently. An aspherical explosion can
produce such a structure of the ejecta. We also study detailed
nucleosynthesis calculations for hyper-energetic supernova explosions
and compare the yields with those of normal supernovae.

\end{abstract}


\section{INTRODUCTION}\label{sec:intro}

On 1998 April 25, GRB980425 triggered the Narrow Field Instrument
(NFI) and Wide Field Camera (WFC) detectors on board {\sl BeppoSAX}
(Soffita et al.  1998). SN~1998bw was discovered within the WFC error
box in the optical (Galama et al. 1998a) and radio wavelength bands
(Kulkarni et al. 1998a) only 0.9 and 3 days after the date of the GRB,
respectively. The X-ray afterglow detected by the NFI within the error
box of GRB980425 is consistent with SN~1998bw (Pian et al. 1999). The
small likelihood of finding a supernova and a GRB in such a small
field over such a short interval of time suggests that SN~1998bw and
GRB980425 are related (Galama et al. 1998b; Kulkarni et al. 1998b).

The early optical spectra of SN 1998bw apparently lack dominant line
features, displaying only some broad emissions at 500, 620 and 800 nm
(Galama et al.  1998b; Stathakis et al. 2000; Patat et al. 2000).
These were shown to be the emission components of P-Cygni profiles of
very broad line blends. They are caused mostly by FeII lines in the
blue, by SiII near 600 nm and by OI + CaII near 720 nm (Iwamoto et al. 
1998, hereafter IMN98). The absence of any hydrogen line after
subtraction of the galaxy background and the fact that the Si II line
at 615 nm has a very large velocity indicate that the supernova is
neither a Type II nor a normal Type Ia (Galama et al. 1998b).
Following the conventional classification scheme (Filippenko 1997),
the lack of strong He I features led Patat \& Piemonte (1998) to
conclude that SN 1998bw is likely to be a Type Ic supernova (SN Ic)
rather than a Type Ib.

The light curve of SN 1998bw (Galama et al. 1998b) showed a very early
rise and reached a peak at $\sim$ 17 days (in the V band) after the
explosion, and then has been declining exponentially with time
(McKenzie \& Schaefer 1999; Patat et al. 2000). This clearly indicates
that the light curve is not a typical optical afterglow of a gamma-ray
burst, but it is powered by the radioactive decay of $^{56}$Ni and
$^{56}$Co as in usual supernovae (IMN98).  The distance modulus to
SN~1998bw is estimated as $\mu = 32.89$~mag, so that the peak
luminosity of SN~1998bw is $\sim 1 \times 10^{43}$ ergs sec$^{-1}$,
which is about ten times brighter than typical SNe Ib/Ic (Clocchiatti
\& Wheeler 1997). To achieve such a high luminosity, a large amount of
$^{56}$Ni must have been synthesized in SN1998bw (IMN98; Woosley,
Eastman \& Schmidt 1999), again about ten times as much as that
produced in typical core-collapse-induced supernovae. The very broad
spectral features and the light curve shape led various groups to the
conclusion that SN 1998bw had a very large {\sl kinetic} energy of
explosion $E_{\rm K}$ (IMN98; Woosley et al. 1999; Branch 2000).

IMN98 constructed models of the core-collapse-induced explosion of C+O
cores of initially massive stars that had lost their hydrogen and
helium-rich layers before the explosion.  Among those models, the
energetic explosion ($E_{\rm K} \sim (2-5) \times 10^{52}$ ergs) of a
C+O star of 13.8 $M_\odot$ successfully fit the first 60 days of the
light curve and the early spectra, although the synthetic spectral
lines were still narrower than the very broad observed features
(IMN98).  Since the kinetic energy is more than one order of magnitude
larger than the energy of typical supernovae, SN 1998bw was called a
``hypernova'' (IMN98), a term we use to describe events with $E_{\rm
K} \gsim 10^{52}$ erg without specifying whether the central engine is
a collapser (MacFadyen \& Woosley 1999), a magnetar (Nakamura 1998;
Wheeler et al. 2000) or a pair-instability supernova.

Interestingly, photometry after $\sim 60$ days showed that SN~1998bw
declined significantly more slowly than the rate predicted by the
model of IMN98 (McKenzie \& Schaefer 1999; Patat et al. 2000). Also
the bolometric light curve has been constructed up to day 500 (Patat
et al. 2000; Sollerman et al. 2000).  Therefore, we have recomputed
the light curve of SN~1998bw using progenitors of different masses and
explosions of different energies, and provide better estimates for
these parameters.

The hydrodynamical models for SN~1998bw are described in \S
\ref{sec:hm}.  Sections \ref{sec:lc} and \ref{sec:spectra} are devoted
to light curve models and synthetic spectra models, respectively.
Explosive nucleosynthesis in SN~1998bw is discussed in
\S\ref{sec:nuc}.  Finally, the nature of this peculiar supernova is
summarized in \S \ref{sec:discussion}, with the emphasis being placed
on possible evidence that the explosion was aspherical.  Preliminary
results have already been reported by Nakamura et al. (1999a, 2000)
and Nomoto et al. (2000ab).

\section{Hydrodynamical models for SN~1998bw}\label{sec:hm}

Hydrodynamical models are constructed as follows.  C+O stars are
chosen as progenitors as in IMN98.  Light curves and spectra are
computed for various C+O star models with different values of the
kinetic energy $E_{\rm K}$ and the ejecta mass $M_{\rm ej}$.  These
parameters can be constrained by comparing calculated light curves,
synthetic spectra, and photospheric velocities with the observational
data of SN 1998bw.

\begin{enumerate}

\item In the ordinary, low energy SN Ic model (model CO138E1), a C+O
star with a mass $M_{\rm CO}= 13.8 M_\odot$ (which is the core of a 40
$M_\odot$ main-sequence star; Nomoto \& Hashimoto 1988; Nomoto et
al. 1997) explodes with $E_{\rm K} = 1.0 \times 10^{51}$ ergs and
$M_{\rm ej} = M_{\rm CO}-M_{\rm cut} \simeq 12 M_\odot$.  $M_{\rm
cut}$ (= 2 $M_\odot$ in this case) denotes the position of the mass
cut, which corresponds to the mass of the compact star remnant. This
is either a neutron star or a black hole, depending on $M_{\rm cut}$.

\item For the hypernova models CO138E50, CO138E30, and CO138E7, the
progenitor C+O star of $M_{\rm CO}= 13.8 M_\odot$ is the same as in
CO138E1. These models have different explosion energies: $E_{\rm K} =
5 \times 10^{52}$ erg (CO138E50), $3 \times 10^{52}$ erg (CO138E30),
and $7 \times 10^{51}$ erg (CO138E7). The ejecta mass is $M_{\rm ej}
\simeq 10-11.5M_\odot$, i.e., $M_{\rm cut} \simeq$ 2.5 - 4 $M_\odot$.
The parameters of the models are summarized in Table \ref{tab:models}. 
The position of the mass cut is chosen so that the ejected mass of
$^{56}$Ni is the value required to explain the observed peak
brightness of SN~1998bw by radioactive decay heating. The compact
remnant in these hypernova models may well be a black hole, because
$M_{\rm cut}$ can exceed, sometimes significantly, the maximum mass of
a stable neutron star.

\end{enumerate}

The hydrodynamics at early phases was calculated using a Lagrangian
PPM code (Colella \& Woodward 1984). All models are spherically
symmetric.  The explosion is triggered by depositing thermal energy in
a couple of zones just below the mass cut so that the final kinetic
energy has the required value. A strong shock wave forms and
propagates toward the surface.  Explosive nucleosynthesis takes place
behind the shock wave.  Radioactive $^{56}$Ni is produced in the deep,
low velocity layers of the ejecta.

In SN II and SN Ib models, it has been demonstrated that
Rayleigh-Taylor instabilities develop at the H/He and He/C+O
interfaces and induce mixing of elements in the ejecta (e.g. Arnett et
al. 1989; Hachisu et al. 1991, 1994; Iwamoto et al. 1993).  Although
bare C+O stars, the progenitors of SNe Ic, lack composition
interfaces with such pronounced density jumps, Rayleigh-Taylor
instabilities can be driven by neutrino heating and
develop at the Ni+Si/O interface (Kifonidis et al. 2000).
In addition, polarization measurements suggested that SNe
Ic, and hypernovae in particular, are asymmetric explosions (Wang et
al. 1999; Danziger et al. 1999; Patat et al.  2000). The asymmetry of
the explosion may reinforce the instability and bring heavy elements
up to high velocity layers. This is particularly true of jet-like
explosions, where it is likely that extensive mixing takes place in
velocity space.  In view of the large uncertainties in our knowledge
of the mixing process, we assume that the material ejected is
uniformly mixed out to a velocity $v = v_{\rm mix}$.

The hydrodynamical models become homologous ($v \propto r$) at $t \sim
250$ sec, and are then used as input for a radiation transfer code.
The lines in Figure \ref{fig:vvsrho} show the density--velocity
distribution of the homologously expanding ejecta (top) and the
enclosed mass $M_r$ as a function of velocity (bottom) for models
CO138E50 (solid), CO138E30 (long-dashed), CO138E7 (short-dashed), and
CO138E1 (dash-dotted).

\section{Light curve models}\label{sec:lc}

Synthetic light curves are computed with a radiative transfer code
(Iwamoto et al. 2000) which takes into account the balance between
photo-ionizations and recombinations and includes a simplified
treatment of line opacity.  The width of the light curve peak,
$\tau_{\rm LC}$, depends on $E_{\rm K}$ and $M_{\rm ej}$ approximately
as $\tau_{\rm LC} \sim (\kappa/c)^{1/2} M_{\rm ej}^{3/4} E_{\rm
K}^{-1/4}$, where \(\kappa\) and \(c\) are the optical opacity and the
speed of light, respectively (Arnett 1982). This means that the light
curve can be reproduced with different explosion models that have the
same values of $M_{\rm ej}^{3} E_{\rm K}^{-1}$.  However, these
parameters can be further constrained from both photospheric
velocities and spectra because the velocity scales roughly as $M_{\rm
ej}^{-1/2} E_{\rm K}^{1/2}$.  The light curve shape depends also on
the distribution of the radioactive heating source $^{56}$Ni, for
which we examine the dependence on $v_{\rm mix}$ (\S\ref{sec:hm}).

Our calculations are compared with the bolometric light curve of SN
1998bw constructed by Patat et al. (2000), which uses a redshift
distance $\mu = 32.89$~mag ($d=37.8$ Mpc, $H_0=65$ km s$^{-1}$
Mpc$^{-1}$) and $A_V = 0.05$~mag. The time of core collapse is set at
the detection of GRB 980425.  Figure \ref{fig:lc2} shows the
bolometric light curves of the energetic models CO138E50 (solid) and
CO138E30 (long-dashed), and Figure \ref{fig:lc3} the less energetic
models CO138E7 (short-dashed) and CO138E1 (dash-dotted).

Figure \ref{fig:vph} shows the evolution of the calculated
photospheric velocities of CO138E50 (solid line), CO138E30
(long-dashed line), CO138E7 (short-dashed line), and CO138E1
(dash-dotted line) compared with those obtained from spectral models
(filled circles), and with the observed velocity of the Si II 635.5 nm
doublet measured in the spectra at the absorption core (open circles;
Patat et al. 2000), and that of the Ca II H + K doublet measured in
the spectrum of May 23 (square; \cite{pp98}).

\subsection {Early Phase}\label{sec:ep}

The early part of the light curve ($t \lsim 25$ days) is well
reproduced by the two energetic models, CO138E50 and CO138E30. The
difference between the two curves is too small to see on the scale of
Figure \ref{fig:lc2} because the light curve width scales as $E_{\rm
K}^{-1/4}$. The early part of the light curve also depends on the
$^{56}$Ni distribution in the ejecta. To reproduce the early sharp
rise of the light curve, extensive mixing of $^{56}$Ni, out to $v_{\rm
mix} = 22,000$ km s$^{-1}$ for CO138E50 and $v_{\rm mix} = 30,000$ km
s$^{-1}$ for CO138E30, is required. A better fit can be obtained if we
adopt the following ad hoc $^{56}$Ni distribution: X($^{56}$Ni) = 0.79
at $v \le 11,000 $ km s$^{-1}$, 0.011 at $ v = $ 11 - 17,000 km
s$^{-1}$, and 0.032 at $v = $ 17 - 40,000 km s$^{-1}$, where
X($^{56}$Ni) denotes the mass fraction of $^{56}$Ni. The result is
shown in Figure \ref{fig:lce50}. Such a non-uniform distribution of
$^{56}$Ni might be due to the mixing caused by Rayleigh-Taylor
instability (\S \ref{sec:hm}), or reflect a complicated structure of a
possible jet-like ejecta (\S \ref{sec:discussion}). The $^{56}$Ni mass
is determined to be 0.4 $M_\odot$ from the fitting of the maximum
luminosity and the date of maximum.  (Note that IMN98 slightly
overestimated the mass of $^{56}$Ni because they assumed that the
bolometric correction is negligible and adopted a different absorption
and distance.)

In contrast, the early light curves of the less energetic models
(CO138E7 and CO138E1) evolve too slowly, reaching the maximum too late
as compared with the observations (Figure \ref{fig:lc3}), even if
$^{56}$Ni is distributed uniformly throughout the ejecta.

The photospheric velocities $v_{\rm ph}$ provide clearer diagnostics
to distinguish between CO138E50 and CO138E30.  In particular, in the
early phase ($t <$ 20 days), $v_{\rm ph}$ of CO138E50 is in good
agreement with the observed $v_{\rm ph}$, while $v_{\rm ph}$ of
CO138E30 is clearly too low to be consistent with the observations
(Fig. \ref{fig:vph}).

In the early phase, therefore, model CO138E50 shows the best agreement
with both the light curve and the photospheric velocities of
SN~1998bw. The synthetic spectra also require the most energetic model
CO138E50 as will be shown in \S\ref{sec:spectra}. CO138E50 has the
same mass as the best model in IMN98 but a larger $E_{\rm K}$, which
is necessary especially to improve the fit to the spectra (\S
\ref{sec:spectra}).

\subsection {Intermediate Phase}

At intermediate phases ($t \sim 25$ - $200$ days), the decline rate of
the light curve is determined mainly by the fraction of the
$\gamma$-rays emitted by $^{56}$Co decay which are trapped in the
ejecta.  The optical depth of the ejecta to the $\gamma$-rays scales
as $\kappa_\gamma \rho R \propto M R^{-2} \propto M^2 E^{-1}_{\rm K}
t^{-2}$. Thus the behaviour of the various models can be seen more
easily than at earlier phases as follows:

\begin{enumerate}

\item The light curve of CO138E50 is consistent with SN1998bw until
day 50 but declines faster than the observation afterwards (Figure
\ref{fig:lc2}; McKenzie \& Schaefer 1999; Patat et al. 2000).  The
light curve of CO138E30 declines more slowly than that of CO138E50,
and is in better agreement with SN1998bw, but it still declines faster
than the observations.

\item The photospheric velocity $v_{\rm ph}$ shows a similar tendency.
Figure \ref{fig:vph} shows that $v_{\rm ph}$ of CO138E50 is the best
fit to the data for the first 20 days, but afterwards $v_{\rm ph}$ of
CO138E30 gives as good a fit. The other two models do not fit $v_{\rm
ph}$ at all.

\item Figure \ref{fig:lc3} shows that the apparently exponential
decline of SN1998bw after day 60 (McKenzie \& Schaefer 1999; Patat et
al. 2000) is well reproduced by the lower energy model CO138E7.  In
this model, $\gamma$-ray trapping is more efficient than CO138E50 and
CO138E30, but the observed low flux level at this phase requires a
reduced $^{56}$Ni mass of $0.28 M_\odot$ (dotted line). Note, however,
that $0.28 M_\odot$ of $^{56}$Ni is too small to reproduce the
observed light curve maximum.  Note also that because of the interplay
of $E_K$ and $M$(Ni), it is difficult to establish those parameters
from the intermediate phase light curve alone.

\item The normal SN Ic model CO138E1 in Figure \ref{fig:lc3} is slow
enough to trap most of the $\gamma$-rays emitted from the $^{56}$Co
decay.  The decline of the light curve is therefore too slow compared
with the observed rate.

\end{enumerate}

These comparisons between SN 1998bw and the model light curves and
$v_{\rm ph}$ in Figures \ref{fig:lc2}, \ref{fig:lc3}, \& \ref{fig:vph}
indicate that $\gamma$-ray deposition after about day 50 in SN~1998bw
is more efficient than predicted by CO138E50 and E30, although these
models give an appropriate description of the early light curve and
spectra.  Higher deposition can be achieved if there exists a
significant amount of low-velocity and high-density material.
Indications of the presence of a low-velocity, high-density region
suggest that the ejecta distribution is not spherically symmetric as
will be discussed in \S \ref{sec:discussion}.  We note that such
indications are also found for SN~1997ef (Iwamoto et al. 2000;
Mazzali, Iwamoto, \& Nomoto 2000a), a lower-energy analogue of
SN~1998bw.

\subsection{Late Phase}

After day $\sim$ 200 the decline of the model light curve becomes
slower, and it approaches the half-life of $^{56}$Co decay around day
400.  At $ t \gsim 400$ days most $\gamma$-rays escape from the
ejecta. The $\gamma$-ray deposition fraction at 400 days is 1\%,
1.5\%, 6\%, and 13\% for CO138E50, E30, E7, and E1, respectively. On
the other hand, kinetic energies of positrons emitted from the decay
of $^{56}$Co are supposed to be fully thermalized because of the
postulated weak magnetic field (e.g., Colgate \& Petchek 1979).
Therefore, positron kinetic energy deposition determines the light
curve at $ t \gsim 400$ days (dotted line in Figure \ref{fig:lc2}).
Here the luminosity by positron deposition is given as
\begin{eqnarray} 
L({\rm Co}, e^+) = 1.4 \times 10^{43} {\rm erg/s}
\hspace{1mm}(M{\rm (Ni)}/M_\odot) \nonumber \\
\times{\rm exp}(-t/111.26{\rm (day)}) \times 0.035,
\label{eq:po}
\end{eqnarray}
where the positron fraction in energy is 3.5 percent (e.g., Axelrod
1980).  (See Cappellaro et al. 1997 and Milne et al. 1999
for the light curves including positron escape.)
If the observed tail should follow the positron-powered
light curve, the $^{56}$Co mass could be determined directly.  Since
positron deposition should occur almost on the spot, this
determination does not depend much on any asphericity of the ejecta.

On June 11, 2000, which corresponds to a SN epoch of $t=$ 778 days,
HST observations detected a point-like source at the position of
SN1998bw (Fynbo et al. 2000).  The observed magnitude (V-mag 25.41
$\pm$ 0.25) is consistent with the prediction of CO138E7 (Figure
\ref{fig:lc2}) but brighter than CO138E50 (Figure \ref{fig:lc3}).  The
observed luminosity would even be higher than CO138E7, if the
bolometric correction was significant.  In any case, these
comparisons suggest that the positron contribution is not dominant yet
around day 800.

\section{Synthetic spectra}\label{sec:spectra}

In Figure \ref{fig:spc} we show the synthetic spectra obtained for the
same 3 epochs fitted in IMN98 (continuous lines) compared with spectra
observed at ESO (bold lines). We used model CO138E50 and computed
synthetic spectra with a Monte Carlo model (Mazzali \& Lucy 1993),
improved with the inclusion of photon branching and a new extended and
improved line list (Lucy 1999; Mazzali 2000).  The synthetic spectra
were computed using the luminosity derived from the light curve, a
distance modulus of $\mu = 32.89$~mag and $A_V = 0.05$. The assumption
of low reddening is supported by the upper limit of 0.1~\AA\ in the
equivalent width of the Na~{\sc i}~D line obtained from
high-resolution spectra (Patat et~al.\ 2000). The observed spectra
used here (solid lines) are the `definitive', fully reduced version of
the same ESO spectra shown in IMN98, and are calibrated with respect
to the V photometry. The residual correction factors for the other
bands are usually very close to 1, but they are $\sim 1.1$ for B in
the May 11 and 23 spectra. Therefore, the new models (i.e. 3 epochs of
CO138E50) have somewhat different parameters than those of IMN98.
Both the luminosities and the photospheric velocities are larger than
in IMN98.  The photospheric velocity is now in better agreement with
the measured velocity of the Si~{\sc ii}\ line (IMN98, Figure 3).

The synthetic spectra clearly improve over those of IMN98, especially
at the earliest epochs.  Absorptions not caused by broad blends of
many lines of moderate strength, such as the Si~{\sc ii}\ feature near
6000\AA\ and, in particular, the O~{\sc i}+Ca~{\sc ii}\ feature
between 7000 and 8000\AA, are now much broader, in significantly
better agreement with the data.  Nevertheless, the blue sides of those
absorptions are still too narrow, indicating that even the new model
CO138E50 may not contain enough mass at the highest velocities.

Therefore we introduced an arbitrary change to the original CO138E50
density structure. Several possibilities were tested, and improved
results were obtained when the density slope was reduced from $\rho
\propto r^{-8}$ to $\rho \propto r^{-6}$ at $v >
30,000$~km~s$^{-1}$. This does not introduce a significant change in
$M_{\rm ej}$, and increases $E_{\rm K}$ by only about 10\%, but it
does increase the density at high velocities, leading to significant
absorption at $v \sim 60,000$~km~s$^{-1}$ in the strongest lines,
especially the Ca~{\sc ii}~IR triplet, extending the absorption
troughs to the blue. The corresponding synthetic spectra are shown as
the dotted lines in Figure \ref{fig:spc}. The effect of the change is
of course largest at the earliest epochs. Although the overall
agreement with the observed spectra is better, several problems
remain, the most severe of which is clearly the excessive strength of
the O~{\sc i}\ line at 7200\AA\ on May 11 and 23.  The composition of
the highest velocity ejecta is dominated by O in our 1D models, and it
is difficult to make that line become weaker. On 23 May, the synthetic
Ca~{\sc ii}~IR triplet matches the weak feature at 8000\AA, which is
first seen on 11 May and which continues to grow until it finally
causes the wavelength of the absorption minimum of the entire broad
feature to shift to $\sim 8200$\AA\ (Patat et~al.\ 2000). This is
rather a peculiar behavior, because on 3 May the O~{\sc i}\ and
Ca~{\sc ii}\ lines had to blend much more to give rise to the observed
broad feature, which then had a minimum at 7000\AA. The synthetic
O~{\sc i}\ line is too strong and too fast. The core of the broad
feature, if it is interpreted as O~{\sc i}\ 7774\AA, indicates a
velocity of 10500 \kms\ on May 11 and 6000 \kms\ on May 23. This is
significantly lower than the corresponding velocity of the model
photosphere. Actually, the definition of a photosphere at very red
wavelengths is not very accurate, because the density of spectral
lines is low and so line opacity does not define a pseudo-continuum in
that region. Therefore, the observations probably indicate that a
large fraction of the O is located at low velocities.

A very flat ($\rho \propto r^{-2}$) density distribution was also used
by Branch (2000) to fit the spectrum of SN~1998bw. This dependence is
however too flat when we use our MC model, because the ionization of,
e.g., Ca~{\sc ii}\ does not fall as steeply as he assumed. On the
other hand, Branch's value of $E_{\rm K}$ ($5 \times 10^{52}$ erg) is
similar to ours, but he quotes a mass of $6 M_{\odot}$ above
7000~km~s$^{-1}$, while in our case the mass above that velocity is as
large as $\sim 10 M_{\odot}$. Such a flat density distribution at high
velocities is also required to fit the spectrum of another hypernova,
SN~1997ef (Mazzali et al. 2000a). This might indicate that the
progenitors of these hypernovae underwent very extensive mass loss;
the outer density structure of the ejecta is flatter than that of
ordinary giants, and comparable to that of a mass-losing star.

Clearly, a definitive solution has not been found yet. It is quite
possible that only by taking into account departures from spherical
symmetry will it be possible to obtain a really accurate fit to the
spectra. Nevertheless, considering the complexity of the problem, our
fits at least demonstrate that a large $E_{\rm K}$ is necessary and
that the O-dominated composition of the SN Ic model yields quite a
reasonable reproduction of the observations.

\section{Explosive Nucleosynthesis}\label{sec:nuc}

We calculated explosive nucleosynthesis using a detailed nuclear
reaction network (Thielemann, Nomoto \& Hashimoto 1996; Nakamura et
al. 1999b). Our calculations are performed in two steps. The first
step is a hydrodynamical simulation of the explosion with a small
nuclear reaction network containing only 13 alpha nuclei ($^{4}$He,
$^{12}$C, $^{16}$O, $^{20}$Ne, $^{24}$Mg, $^{28}$Si, $^{32}$S,
$^{36}$Ar, $^{40}$Ca, $^{44}$Ti, $^{48}$Cr, $^{52}$Fe, and $^{56}$Ni),
as described in \S \ref{sec:hm}. In the second step, post-processing
calculations are performed at each mesh point of the hydrodynamical
model with the extended reaction network (Hix \& Thielemann 1996),
which contains 211 isotopes up to $^{71}$Ge.

The top panel of Figure \ref{fig:nssph} shows the isotopic composition
of the ejecta of the hypernova model CO138E50 as a function of the
enclosed mass $M_r$ and of the expansion velocity. The nucleosynthesis
in other hypernova and supernova models (CO138E30, CO138E7, CO138E1)
is also shown in Figure \ref{fig:nssph} for comparison. The yields of
the hypernova and supernova models are summarized in Table
\ref{tab:ns}. Table \ref{tab:ele50} and Figure \ref{fig:ratioE50} give
respectively more detailed yields and the abundances of stable
isotopes relative to the solar values for model CO138E50. From these
figures and tables, we note the following characteristics of
nucleosynthesis of hyper-energetic explosions compared with normal
energy explosions.

\begin{enumerate} 

\item The complete Si-burning region where $^{56}$Ni is produced is
extended further out (in mass coordinates) as the explosion energy
increases. How much processed matter is ejected from this region
depends on the mass cut. Compared with normal core-collapse
supernovae, the much larger amount of $^{56}$Ni ($\sim 0.4M_\odot$)
observed in SN1998bw implies that the mass cut is deeper, so that the
elements synthesized in this region, such as $^{59}$Cu, $^{63}$Zn, and
$^{64}$Ge (which decay into $^{59}$Co, $^{63}$Cu, and $^{64}$Zn,
respectively), are ejected more abundantly. Among hypernova models,
more energetic models produce more $^{56}$Ni in the incomplete Si
burning region (See (3) below), and thus $M_{\rm cut}$ is larger to
eject $\sim 0.4 M_\odot$ $^{56}$Ni as constrained from the light curve
of SN~1998bw (Figure \ref{fig:nssph}).

\item In the complete Si-burning region of the hypernova models
(CO138E50/E30), elements produced by $\alpha$-rich freezeout are
enhanced because nucleosynthesis proceeds at lower densities than in
CO138E1. Figure \ref{fig:nssph} clearly shows a trend that a larger
amount of $^{4}$He is left in more energetic explosions.  Hence, the
mass fractions of the species synthesized through $\alpha$-particle
capture, such as $^{44}$Ti and $^{48}$Cr (which decay into $^{44}$Ca
and $^{48}$Ti, respectively) is larger in CO138E50/E30/E7 than
CO138E1. The integrated mass of these species depends on $M_{\rm
cut}$. For CO138E50, the ejected mass of $^{44}$Ti is smaller than in
other models because of its larger $M_{\rm cut}$. Note that the
$^{4}$He produced even in the most energetic models has a velocity
significantly smaller than that of the He shell identified in
SN~1998bw at a velocity of 18300\kms\ by Patat et al. (2000) on the
basis of near-IR spectra. Spectral evidence for low velocity He in
SN~1998bw is unclear, however the distribution of He depends on mixing
in velocity space.

\item The incomplete Si-burning region is more massive in more
energetic explosions.  The main products in this region are $^{28}$Si,
$^{32}$S, and $^{56}$Ni. CO138E50 produces 0.2$M_\odot$ of $^{56}$Ni
in this region. Other important species such as $^{52}$Fe, $^{55}$Co,
and $^{51}$Mn (decaying into $^{52}$Cr, $^{55}$Mn, and $^{51}$V,
respectively) are synthesized more abundantly in the more energetic
explosions.

\item For the larger explosion energy, oxygen burning takes place in
more extended, lower density regions. O, C, and Al are burned more
efficiently in these cases, and the abundances of the elements in the
ejecta are smaller, while a larger amount of ash products such as Si,
S and Ar is synthesized by oxygen burning.

\end{enumerate}

\section{Conclusions and Discussion}\label{sec:discussion}

In this paper, we have presented a model for SN~1998bw which is in
better agreement with the early observations than the previous model
in IMN98. Models with different $E_{\rm K}$ yield different synthetic
spectra, and by comparing with the observed early-time spectra of
SN~1998bw and trying to fit the very broad absorption features, we
selected model CO138E50 with $M_{\rm ej} = 10 M_{\odot}$ and $E_{\rm
K} = 5 \times 10^{52}$~erg as the best match to the early data. The
large value of $E_{\rm K}$ qualifies SN~1998bw as `the' Type Ic
Hypernova. (SN IIn 1997cy may be called a ``Type IIn Hypernova'';
Germany et al. 2000; Turatto et al. 2000).  The mass of the progenitor
C+O star is $13.8 M_{\odot}$, corresponding to a main sequence mass of
$\sim 40 M_{\odot}$. All models require $M$($^{56}$Ni)$ \sim 0.4
M_{\odot}$ to power the bright light curve peak \footnote{ After
submitting this paper, Sollerman et al.'s paper (2000) was published.
They found that $^{56}$Ni of $\sim 0.3$ - $0.9 M_\odot$ was necessary
to power the late light curve based on the model CO138 (IMN98), which
is consistent with our results.}.  This is about an order of magnitude
larger than in typical core-collapse SNe. The compact remnant is
probably a black hole, because its mass exceeds $\sim 3 M_\odot$ as
constrained from the mass of $^{56}$Ni.

Although the early light curve of SN1998bw ($t \lsim$ 50 days) is
reproduced well by the most energetic model CO138E50, the observed
tail declines more slowly than this model does (\S \ref{sec:lc}).  The
lower energy model CO138E30 is in better agreement with observations
at $t \lsim$ 100 days but it still declines is too fast.  Model
CO138E7 has a slower tail and can reproduce the observed light curve
tail if a smaller $^{56}$Ni mass of 0.28 $M_{\odot}$ is adopted. This
suggests that there might be a high density core with low velocities
in SN1998bw where the $\gamma$-rays deposit efficiently.

Such a dense core could be formed in spherically symmetric models if
the exploding star had a massive He envelope so that a strong reverse
shock was formed at the C+O/He interface and largely decelerated the
inner core as found in SN Ib models (Hachisu et al. 1991, 1994).
However, the apparent absence of the He I lines in the optical spectra
in SN 1998bw (Patat et al. 2000) is not consistent with the presence
of such a massive He envelope.  Also late time energy input from
$^{56}$Co bubbles might create a non-homologous structure.  However,
the energy from radioactive decays is much smaller than the large
explosion energy of SN~1998bw ($\sim 10^{52}$ ergs), so that its
effect is negligible.

We suggest that the peculiar density distribution is the result of a
non-spherically symmetric explosion.  If the outburst in SN1998bw took
the form of a prolate spheroid, for example, the explosive shock was
probably strong along the long axis, ejecting material with large
velocities and producing abundant $^{56}$Ni, which might have caused
the early bright light curve. In directions away from the long axis,
on the other hand, oxygen would not be consumed, and the density could
be high enough for $\gamma$-rays to be trapped even at advanced
phases, thus giving rise to the slowly declining tail. These features
can be seen in the hydrodynamical models of jet-like explosions
(Maeda et al. 2000; MacFadyen \& Woosley 1999; Khokhlov et al. 1999; 
Nagataki et al. 1997). The fact that the late light curve is fitted by
model CO138E7 with 0.28 $M_\odot$ of $^{56}$Ni, while the early light
curve requires as much as 0.4 $M_\odot$, could also support this
suggestion, and point at the geometrical ratio between the extension
of the high- and low-density regions.

There are several observations which support the above non-spherical
explosion scenario.  The observed polarization ($\sim 0.5$\%) of the
early optical light (Patat et al. 2000; IMN98) suggests that the
ejecta of SN~1998bw is aspherical. The
observed abnormal distribution of elements in velocity space as seen
in the emission line profiles in the nebular phase, with significant amounts
of O being located at lower velocity than Fe (Patat et al. 2000;
Danziger et al. 1999; Nomoto et al. 2000) can better be explained if
the explosion is non-spherically symmetric (Maeda et al. 2000).

The possible connection between SN~1998bw and GRB980425 also supports
the conjecture that SN~1998bw was aspherical. The energy of the photons
produced by synchrotron emission at the relativistic shock is
approximately given by $h\nu \sim 160$ ${\rm keV}$ $(\Gamma / 100)^4
n_1^{1/2}$ (Piran 1999), where $\Gamma$ is the Lorentz factor of the
shock and $n_1$ (cm$^{-3}$) is the density of the interstellar matter. 
In order to produce an observable GRB, $\Gamma$ should be as large as
$\Gamma \sim 100$. However, even the most energetic model (CO138E50)
has only a very small mass of relativistic ejecta ($\sim 10^{-10}
M_\odot$ with $\Gamma \ge 100$), although relativistic hydrodynamical
calculations are necessary to obtain the accurate mass.  Such a small
amount of relativistic material, which is consistent with previous
estimates (IMN98; Woosley et al. 1999), is not enough to produce
GRB980425 in a spherically symmetric model.  However, if the explosion
is axi-symmetric, for instance, the energy can be carried by only a
small fraction of the material, which might then attain a large
Lorentz factor.

Regarding asphericity, we note that Danziger et al. (1999) and Mazzali
et al.  (2000b; see also Nomoto et al. 2000a) estimated a $^{56}$Ni
mass of 0.35 - 0.65 $M_\odot$ from the nebular lines of Fe of
SN~1998bw. This estimate does not depend much on the asphericity, and
is in good agreement with the $^{56}$Ni mass of the spherical models
CO138. On the other hand, H\"oflich et al. (1999) suggested that the
$^{56}$Ni mass can be as small as 0.2 $M_\odot$ if aspherical effects
are large. Because the difference between these results is not so
large, aspherical effects might be modest in SN~1998bw. If the
$^{56}$Ni mass could be determined more accurately from the late
observations, it would provide a good measure of the degree of the
asphericity.

\bigskip

We would like to thank Ferdinando Patat and Enrico Cappellaro for
useful information and discussion on the observations of SN1998bw.
This work has been supported by the grant-in-Aid for Scientific
Research (12640233, 12740122) and COE research (07CE2002) of the
Japanese Ministry of Education, Science, Culture, and Sports.

\clearpage

\begin{figure}
\epsscale{.45}
\plotone{fig1a.epsi}
\end{figure}
\begin{figure}
\epsscale{.45}
\plotone{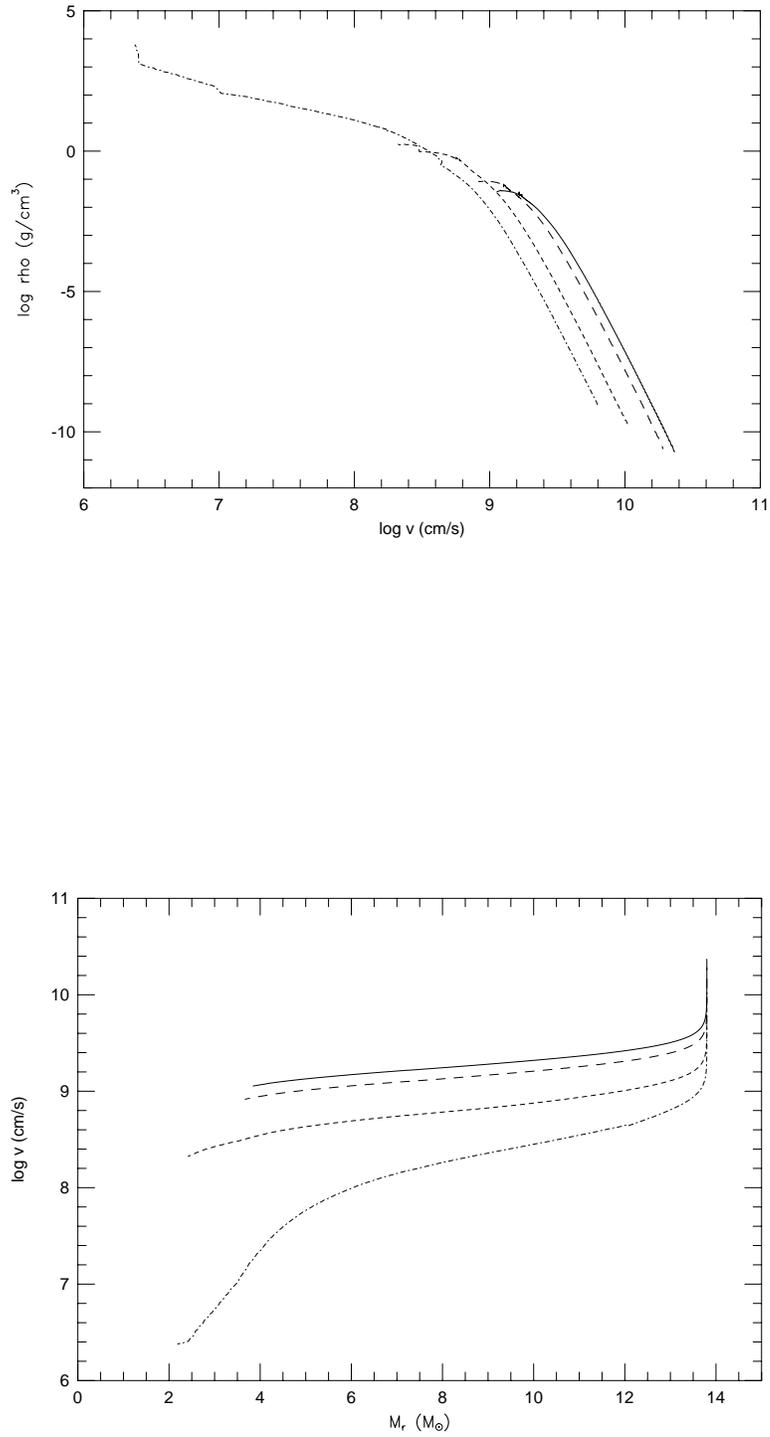}
\caption{
Density distributions against the velocity of homologously
expanding ejecta (above)
and the velocity profiles ageinst the enclosed mass (below)
for CO138E50 (solid line),
CO138E30 (long-dashed line), CO138E7 (short-dashed line),
and CO138E1 (dash-dotted line) at $t =$ 250 sec.
\label{fig:vvsrho}}
\end{figure}

\begin{figure}
\epsscale{.7}
\plotone{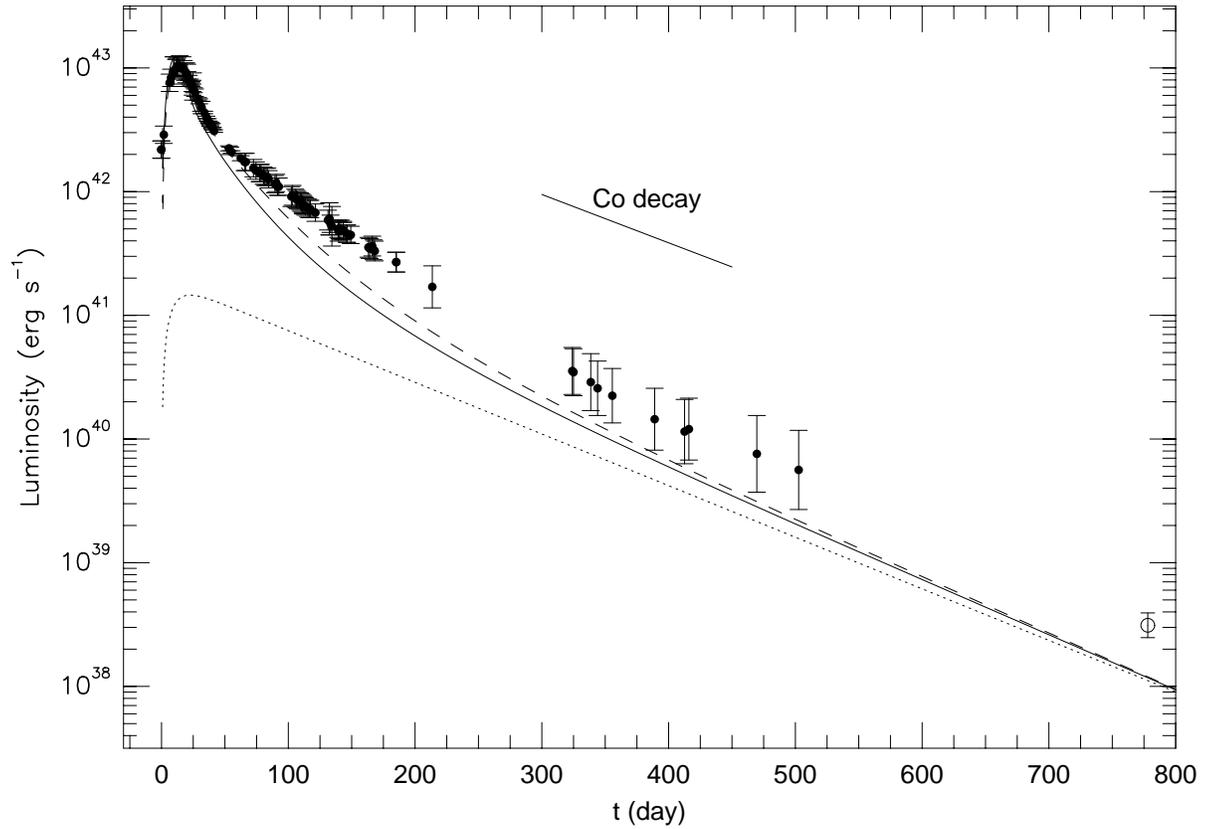}
\caption{The light curves of CO138E50 ($E_{\rm K} = 5 \times 10^{52}$
erg; solid line) and CO138E30 ($E_{\rm K} = 3 \times 10^{52}$ erg;
long-dashed line) compared with the bolometric light curve of SN1998bw
(Patat et al. 2000).  A distance modulus of $\mu = 32.89$ mag and $A_V
= 0.05$ are adopted.  The dotted line indicates the energy deposited
by positrons for CO138E50.  The HST observation at day 778 (Fynbo et
al. 2000) is shown by assuming negligible bolometric correction.
\label{fig:lc2}}
\end{figure}

\begin{figure}
\epsscale{.7}
\plotone{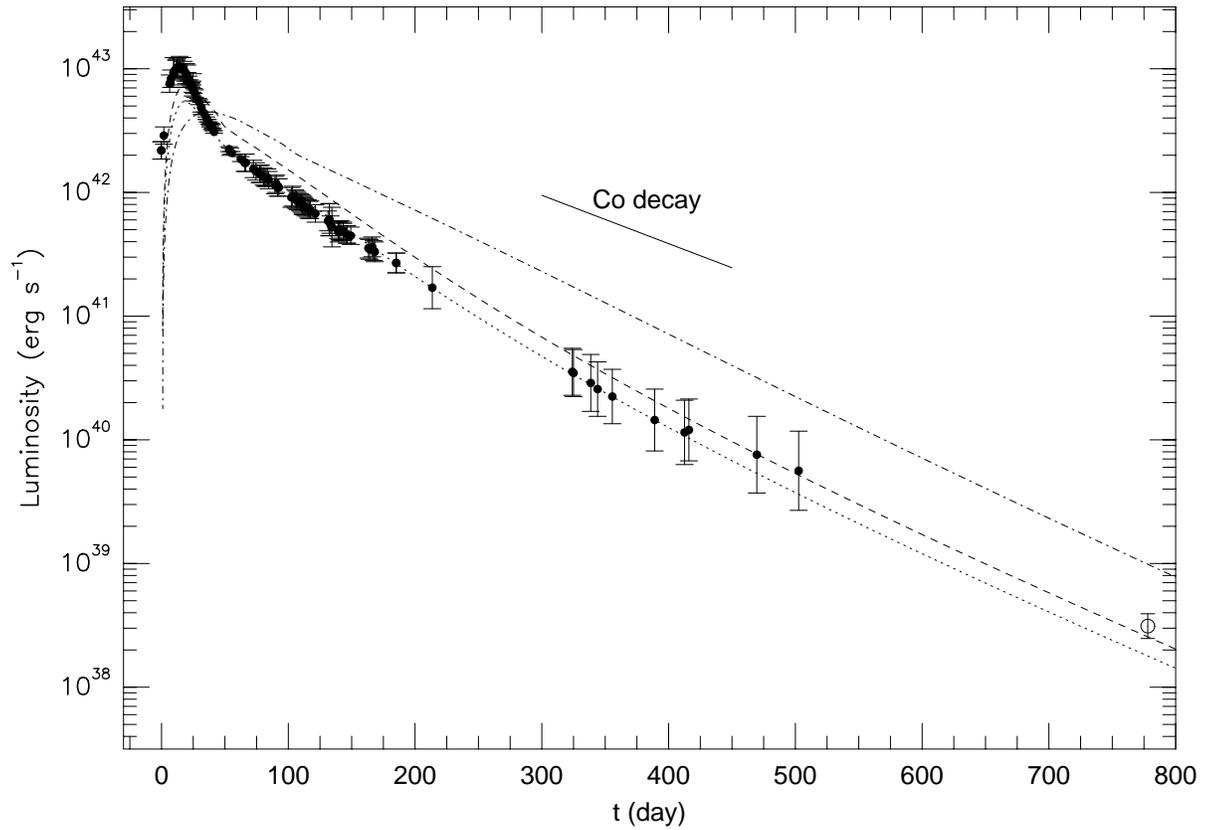}
\caption{The light curves of CO138E7 ($E_{\rm K} = 7 \times 10^{51}$
erg; dashed line) and CO138E1 ($E_{\rm K} = 1 \times 10^{51}$ erg:
dash-dotted line) compared with the bolometric light curve of SN1998bw
(Patat et al. 2000).  Also shown is the light curve of modified
CO138E7 with smaller $^{56}$Ni mass of 0.28$M_\odot$ (dotted line).
The HST observation at day 778 (Fynbo et al. 2000) is shown by
assuming negligible bolometric correction.
\label{fig:lc3}}
\end{figure}

\begin{figure}
\epsscale{.7}
\plotone{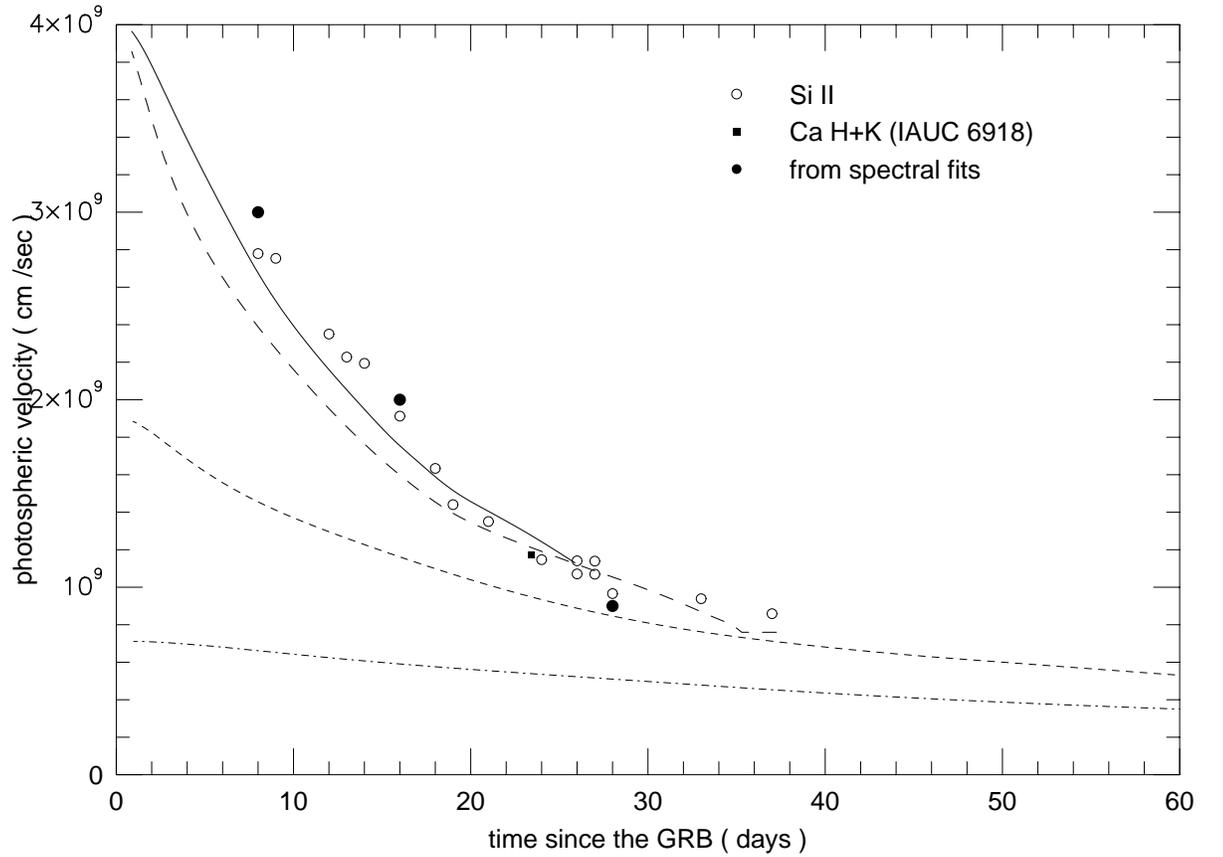}
\caption{Photospheric velocities of
CO138E50 (solid line), CO138E30 (long-dashed line),
CO138E7 (short-dashed line), and CO138E1 (dash-dotted line)
compared with the observations of SN1998bw.
Shown are the photospheric velocities obtained
from spectral models (filled circles, this paper),
the observed velocity of the Si II 634.7, 637.1 nm lines
measured in the spectra at the absorption core
(open circles; Patat et al. 2000),
and that of the Ca II H + K doublet measured in the spectrum
of May 23 (square; Patat \& Piemonte 1998).
\label{fig:vph}}
\end{figure}

\begin{figure}
\epsscale{.7}
\plotone{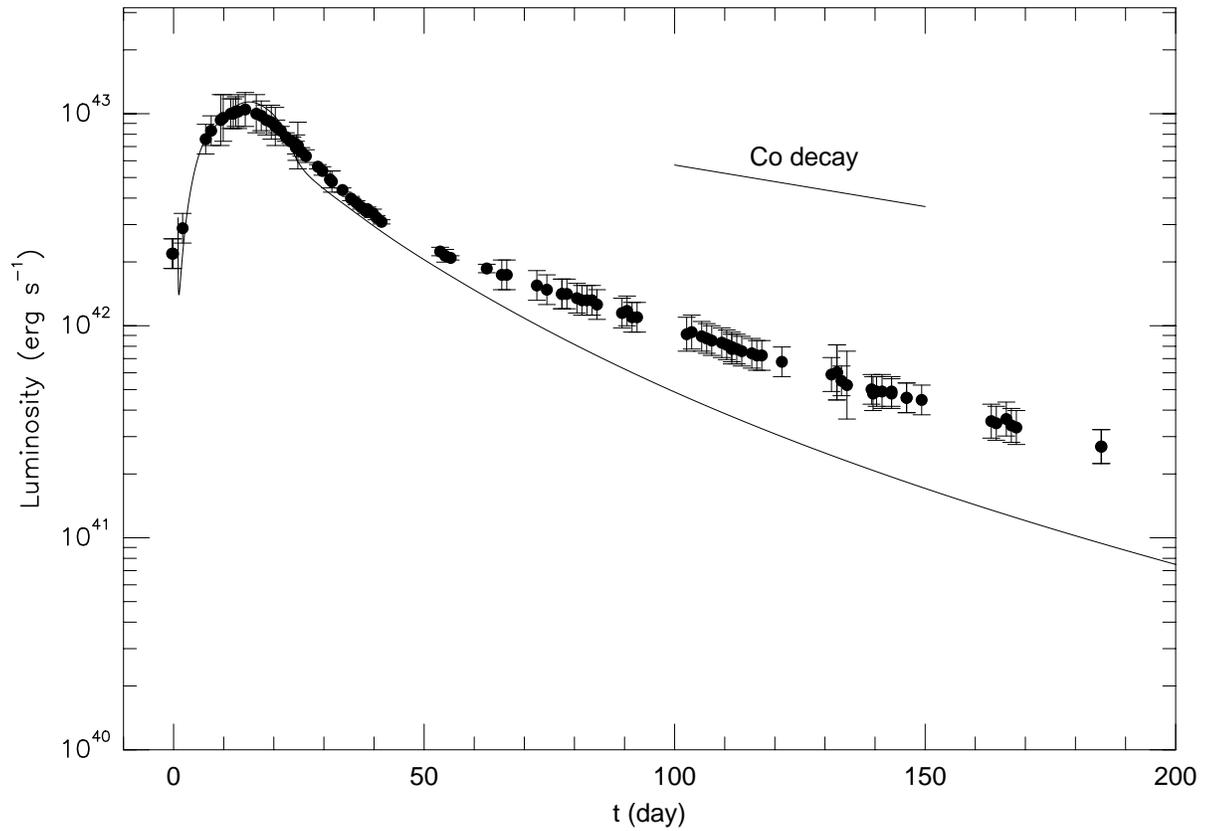}
\caption{The eraly light curves of model CO138E50
with modified $^{56}$Ni distribution (see text)
compared with the bolometric light curve of SN1998bw (Patat et al. 2000).
\label{fig:lce50}}
\end{figure}

\begin{figure}
\epsscale{.7}
\plotone{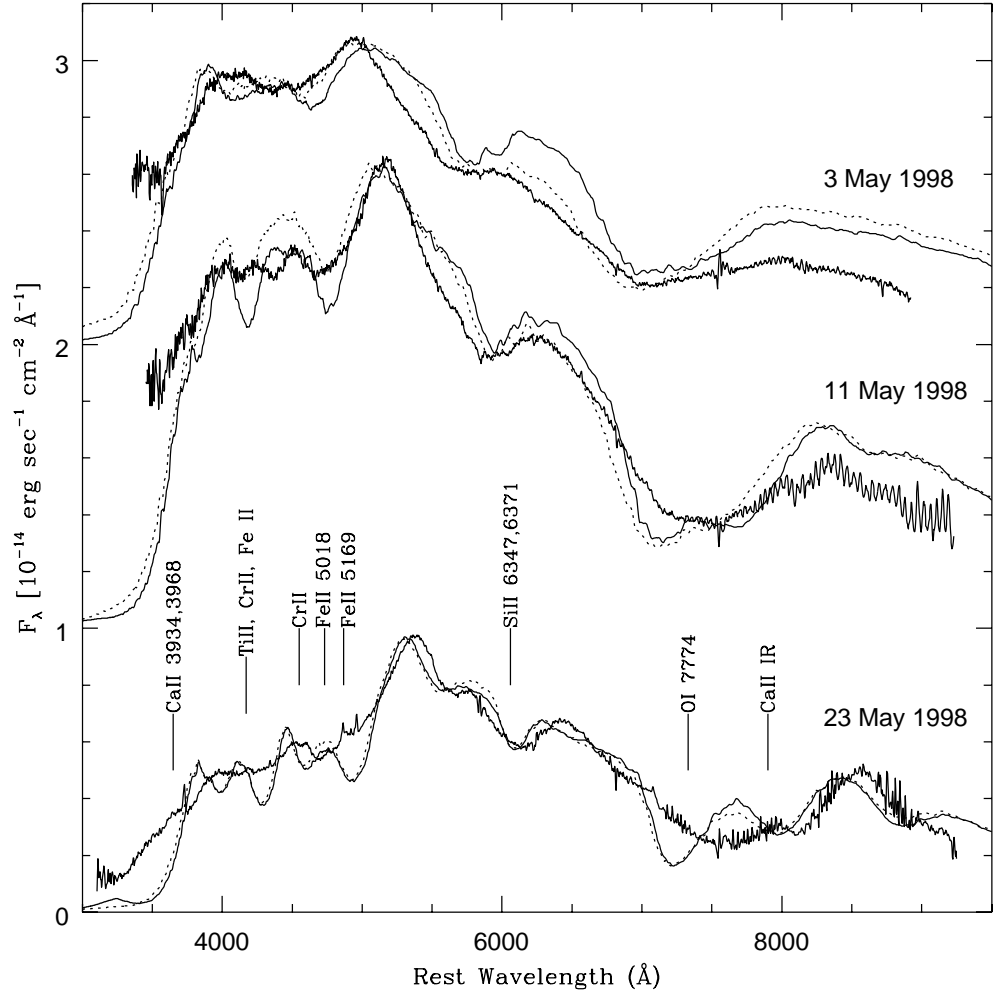}
\caption{Three ESO spectra of SN~1998bw near maximum (bold lines) are
compared with synthetic spectra obtained using model CO138E50
(continuous lines)
and with spectra obtained for a modified density distribution as described
in the text (dotted lines).
\label{fig:spc}}
\end{figure}

\clearpage

\begin{figure}
\epsscale{.5}
\plotone{fig7a.epsi}
\end{figure}
\begin{figure}
\epsscale{.5}
\plotone{fig7b.epsi}
\end{figure}
\begin{figure}
\epsscale{.5}
\plotone{fig7c.epsi}
\end{figure}
\begin{figure}
\epsscale{.5}
\plotone{fig7d.epsi}
\caption{
The isotopic composition of ejecta of hypernovae
($E_{\rm K} = 5 \times 10^{52}$ ergs; top left),
($E_{\rm K} = 3 \times 10^{52}$ ergs; top right),
($E_{\rm K} = 7 \times 10^{51}$ ergs; bottom left),
and usual supernovae
($E_{\rm K} = 1 \times 10^{51}$ erg; bottom right).
Only the dominant species are plotted.
The explosive nucleosynthesis is calculated
using a detailed nuclear reaction network including a total of 211
isotopes up to $^{71}$Ge.
\label{fig:nssph}}
\end{figure}

\begin{figure}
\epsscale{.7}
\plotone{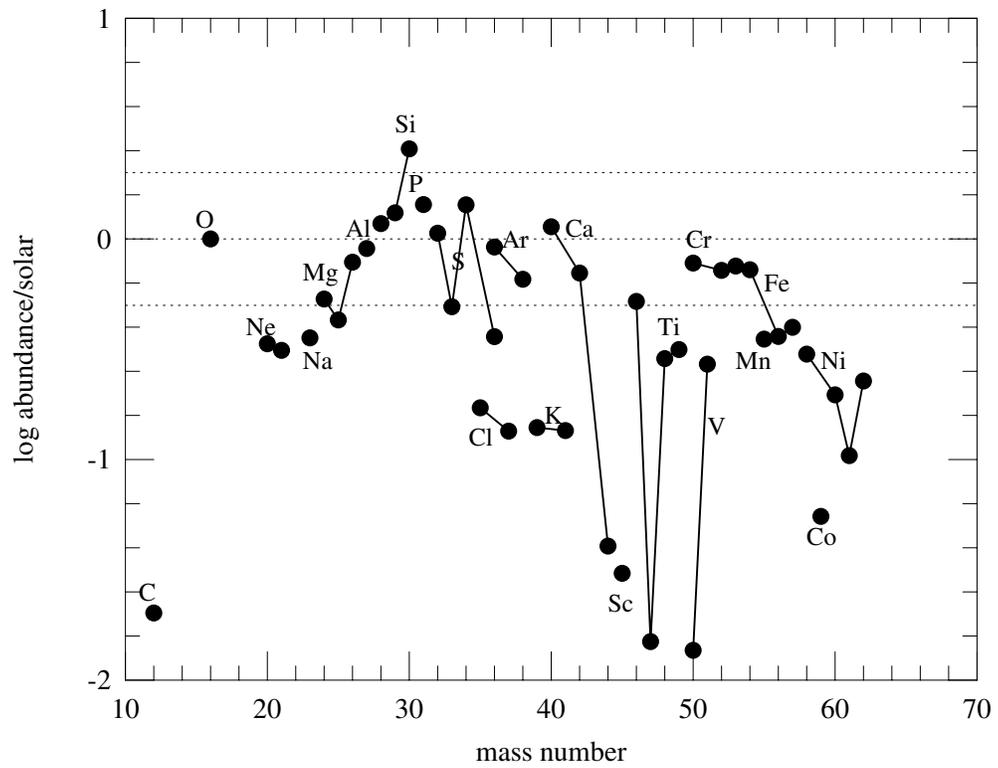}
\caption{
Abundances of stable isotopes
relative to the solar values for model CO138E50.
\label{fig:ratioE50}}
\end{figure}

\clearpage

\begin{table*}
\begin{center}
\begin{tabular}{ccccccc}
\hline model & $M_{\rm ms}${\small($M_\odot$)}
& $M_{\rm C+O}$ & $M_{\rm ej}$ &
 $^{56}$Ni mass & $M_{\rm cut}$
& $E_{\rm K}$ {\small (10$^{51}$ erg)} \\
\hline CO138E1  & $\sim$ 40 & 13.8 & 12   & 0.4 &  2   &  1 \\
\hline CO138E7 & $\sim$ 40 & 13.8  & 11.5 & 0.4 &  2.5 &  7 \\
\hline CO138E30 & $\sim$ 40 & 13.8 & 10.5 & 0.4 &  3.5 & 30 \\
\hline CO138E50 & $\sim$ 40 & 13.8 & 10   & 0.4 &  4   & 50 \\
\hline
\end{tabular}
\end{center}

\tablenum{1}
\caption{\hspace{6.5cm}Parameters of the C+O star models. \label{tab:models}}

\end{table*}

\begin{table*}
\begin{center}
\begin{tabular}{ccccccccccc} \hline
model   & C    &  O  & Ne   & Mg  & Si   & S    & Ca    & Ti    & Fe  &
Ni\\\
CO138E1 & 0.10 & 8.9 & 0.66 & 0.51& 0.52 & 0.19 & 0.023 &0.0007 &0.44
&0.065\\\
CO138E7& 0.09 & 8.5 & 0.58 & 0.47& 0.52 & 0.19 & 0.029 &0.0010 &0.44
&0.036\\
CO138E30& 0.07 & 7.6 & 0.44 & 0.38& 0.74 & 0.35 & 0.054 &0.0009 &0.45
&0.023\\
CO138E50& 0.06 & 7.1 & 0.37 & 0.34& 0.83 & 0.41 & 0.066 &0.0007&0.46
&0.018\\
\hline\hline
model   & $^{44}$Ti           &$^{56}$Ni& $^{57}$Ni \\
CO138E1 & 3.4$\times 10^{-4}$ & 0.40 & 1.6$\times 10^{-2}$ \\
CO138E7& 4.5$\times 10^{-4}$ & 0.40 & 1.4$\times 10^{-2}$ \\
CO138E30& 2.3$\times 10^{-4}$ & 0.40 & 1.2$\times 10^{-2}$ \\
CO138E50& 5.5$\times 10^{-5}$ & 0.40 & 1.1$\times 10^{-2}$ \\
\hline
\end{tabular}
\end{center}

\tablenum{2}
\caption{\hspace{5.5cm}Yields of hypernova and supernova models ($M_\odot$).\label{tab:ns}}

\end{table*}

\begin{table*}
\begin{center}
\begin{tabular}{cccccccccccccccc} \hline
$^{12}$C  & 6.45E-02 & & $^{13}$C  & 2.57E-08 & & $^{14}$N  & 1.34E-07 & &
$^{15}$N  & 4.08E-08 & & $^{16}$O  &   7.07   & \\
$^{17}$O  & 1.76E-07 & & $^{18}$O  & 2.45E-08 & & $^{19}$F  & 1.12E-08 & &
$^{20}$Ne & 3.72E-01 & & $^{21}$Ne & 1.10E-03 & \\
$^{22}$Ne & 5.49E-04 & & $^{23}$Na & 1.18E-02 & & $^{24}$Mg & 2.57E-01 & &
$^{25}$Mg & 2.76E-02 & & $^{26}$Mg & 5.78E-02 & \\
$^{27}$Al & 4.99E-02 & & $^{28}$Si & 7.27E-01 & & $^{29}$Si & 4.30E-02 & &
$^{30}$Si & 5.70E-02 & & $^{31}$P  & 6.93E-03 & \\
$^{32}$S  & 3.88E-01 & & $^{33}$S  & 1.48E-03 & & $^{34}$S  & 2.46E-02 & &
$^{36}$S  & 2.12E-05 & & $^{35}$Cl & 5.17E-04 & \\
$^{37}$Cl & 1.39E-04 & & $^{36}$Ar & 7.10E-02 & & $^{38}$Ar & 1.00E-02 & &
$^{40}$Ar & 2.85E-07 & & $^{39}$K  & 4.26E-04 & \\
$^{41}$K  & 3.21E-05 & & $^{40}$Ca & 6.61E-02 & & $^{42}$Ca & 2.83E-04 & &
$^{43}$Ca & 4.60E-07 & & $^{44}$Ca & 5.53E-05 & \\
$^{46}$Ca & 5.53E-10 & & $^{48}$Ca & 7.90E-14 & & $^{45}$Sc & 1.02E-06 & &
$^{46}$Ti & 1.09E-04 & & $^{47}$Ti & 2.96E-06 & \\
$^{48}$Ti & 5.87E-04 & & $^{49}$Ti & 4.90E-05 & & $^{50}$Ti & 4.15E-09 & &
$^{50}$V  & 1.00E-08 & & $^{51}$V  & 8.39E-05 & \\
$^{50}$Cr & 5.11E-04 & & $^{52}$Cr & 9.53E-03 & & $^{53}$Cr & 1.16E-03 & &
$^{54}$Cr & 3.07E-07 & & $^{55}$Mn & 4.32E-03 & \\
$^{54}$Fe & 4.92E-02 & & $^{56}$Fe & 4.01E-01 & & $^{57}$Fe & 1.07E-02 & &
$^{58}$Fe & 1.04E-07 & & $^{59}$Co & 1.72E-04 & \\
$^{58}$Ni & 1.36E-02 & & $^{60}$Ni & 3.54E-03 & & $^{61}$Ni & 8.69E-05 & &
$^{62}$Ni & 5.91E-04 & & $^{64}$Ni & 3.10E-13 & \\
$^{63}$Cu & 3.47E-07 & & $^{65}$Cu & 2.36E-07 & & $^{64}$Zn & 6.49E-06 & &
$^{66}$Zn & 3.91E-06 & & $^{67}$Zn & 3.03E-09 & \\
$^{68}$Zn & 1.93E-09 & &           &          & &           &          & &
          &          & &           &          & \\\hline
\end{tabular}
\end{center}

\tablenum{3}
\caption{\hspace{6cm}Detaied yields of model CO138E50 ($M_\odot$). \label{tab:ele50}}

\end{table*}

\end{document}